\documentclass[a4paper]{jpconf}

\usepackage{amssymb}
\usepackage{amsmath}
\usepackage{graphicx}
\usepackage{iopams}

\begin{document}

\title{Extreme scenarios of new physics in the UHE astrophysical neutrino flavour ratios}

\author{M Bustamante$^1$, A M Gago$^1$, and C Pe\~na-Garay$^2$}

\address{$^1$ Seccion Fisica, Departamento de Ciencias, Pontificia Universidad Catolica del Peru,
Apartado 1761, Lima, Peru}
\address{$^2$ Instituto de Fisica Corpuscular (IFIC), Centro Mixto CSIC-UVEG Edificio Investigacion Paterna, Apartado 22085, 46071 Valencia, Spain}

\ead{mbustamante@pucp.edu.pe}

\begin{abstract}

We add an energy-independent Hamiltonian to the standard flavour oscillation one. This kind of physics might appear in theories where neutrinos couple differently to a plausible non-zero torsion of the gravitational field or more dramatically in the presence of CPT-violating physics in the flavour oscillations. If this contribution exists, experiments at higher energies are more sensitive to their free parameters, and flavour conversion could be severely modified. We show that this new physics modifies the neutrino mixing angles and find expressions that relate the new, effective, angles to the standard oscillation parameters $\Delta m_{ij}^2$, $\theta_{ij}$ and $\delta_{CP}$ and to the parameters in the new-physics Hamiltonian, within a three-neutrino formalism. We consider scenarios where the new parameters allow for extreme deviations of the expected neutrino flavour ratios at Earth from their standard values. We show that large departures of the standard flavour scenario are plausible, which would be a strong hint of the violation of a conserved symmetry.

\end{abstract}

\section{Motivation and problem}\label{Section_Intro}

By now it has been demonstrated that neutrinos can change flavour. Experiments with solar, atmospheric, reactor and accelerator neutrinos have established that there is a nonzero probability that a neutrino created with a certain flavour is detected with a different one after having propagated for some distance, and that this probability is dependent on the propagated distance, $L$, and the neutrino energy $E$. 

The standard mechanism that explains neutrino flavour transitions makes use of two different bases: the basis of neutrino mass eigenstates, which have well-defined masses, and the basis of neutrino interaction states \--the flavour basis\-- which are the ones that take part in weak processes such as $W$ decay. The two bases are connected by means of a unitary transformation $U$, so that we can write each of the flavour states $\lvert\nu_\alpha\rangle$ as a linear combination of the mass eigenstates $\lvert\nu_i\rangle$, i.e.
\begin{equation}
 \lvert\nu_\alpha\rangle = \sum_{i} U_{\alpha i} \lvert\nu_i\rangle ~,
\end{equation}
where the coefficients $U_{\alpha i}$ are components of the unitary mixing matrix that represents the transformation. Assuming the existence of three active neutrino families ($\alpha = e,\mu,\tau$), as indicated by experiment \cite{ALEPH89}, the summation index runs $i=1,2,3$ and $U$ is a $3\times3$ matrix. Using this definition for the flavour states it is straightforward to conclude that flavour transitions occur because neutrinos are massive, and because different mass eigenstates have different masses. This can be seen, within the three-neutrino formalism, by writing the transition probabilities from $\nu_\alpha$ to $\nu_\beta$ as
\begin{equation}\label{EqTransProb}
 P_{\alpha\beta} 
 = \delta_{\alpha\beta} - 4\sum_{i>j} \text{Re}\left(J_{\alpha\beta}^{ij}\right) \sin^2\left(\frac{\Delta m_{ij}^2}{4E}L\right)
 + 2 \sum_{i>j} \text{Im}\left(J_{\alpha\beta}^{ij}\right) \sin\left(\frac{\Delta m_{ij}^2}{2E}L\right) ~,
\end{equation}
where $\Delta m_{ij}^2 \equiv m_i^2 - m_j^2$, with $m_i$ the mass of the $i$-th eigenstate, and $J_{\alpha\beta}^{ij} \equiv U_{\alpha i}^\ast U_{\beta i} U_{\alpha j} U_{\beta j}^\ast$. Clearly, if $\Delta m_{ij}^2 = 0$, no transitions occur. Note the $1/E$ dependence on the energy associated with this standard, mass-driven, oscillation mechanism.

The experiments that have studied neutrino flavour transitions \cite{Walter08} have been designed to detect neutrinos with energies that range from a few MeV (solar neutrinos) to the TeV scale (atmospheric neutrinos). The mass-driven mechanism mentioned in the preceding paragraph has been experimentally confirmed within this energy range. Notably, data from the Super-Kamiokande atmospheric neutrino experiment \cite{Fogli99} was used to find an energy dependence of the oscillation probability of $E^n$, with $n = -0.9 \pm 0.4$ at $90\%$ confidence level, thus confirming the dominance of the mass-driven mechanism in this energy range, and relegating any other possible mechanism due to new physics to subdominance. It is possible, however, that one or more of such subdominant mechanisms become important at higher energies. 

In the present paper, we have explored a possible scenario where there is an extra oscillation mechanism present which results in an energy-independent contribution to neutrino oscillations. This mechanism, though subdominant in the MeV-TeV range, might become dominant at higher energies. We will deferr the detailed treatment of how the contribution is introduced to Section \ref{Section_Theory} and focus now on the possible mechanisms behind it. 

An energy-independent contribution to neutrino oscillations is introduced by a vector coupling of the form
\begin{equation}\label{EqLagrangian}
 \mathcal{L} = \overline{\nu}^\alpha b_\mu^{\alpha\beta} \gamma^\mu \nu^\beta ~.
\end{equation}
Such a term results in an energy-independent oscillation phase $b_{ij} \equiv b_i - b_j$, with the $b_i$ eigenvalues of the $b$ matrix, so that the phase in Eq.~(\ref{EqTransProb}) becomes
\begin{equation}
 \frac{\Delta m_{ij}^2}{2E} \rightarrow \frac{\Delta m_{ij}^2}{2E} + b_{ij} ~,
\end{equation}
and the mixing matrix $U$ is modified as well. Specially relevant to the detection prospects of new physics is the fact that not only the oscillation phase \--which, as we will see, is lost when considering neutrinos from distant astrophysical sources\--, but also its amplitude, is modified by such an energy-independent contribution. 

A vector coupling like that of Eq.~(\ref{EqLagrangian}) could be induced by violation of Lorentz invariance which in turn results in a violation of CPT symmetry. A realisation of this possibility is the Standard Model Extension \cite{Colladay98, Kostelecky04}, which includes CPT-odd terms in the Lagrangian, such as that of Eq.~(\ref{EqLagrangian}). A different possible mechanism is a nonuniversal coupling to a torsion field \cite{DeSabbata81}.

In the next Section, we will argue that a possible energy-independent contribution to flavour oscillations could become important, with respect to the standard oscillation terms, when the neutrino energy is high (PeV or more). The highest-energy available flux of neutrinos is the expected ultra-high-energy (UHE) flux from astrophysical sources \--notably, active galaxies\-- which are located at distances in the order of tens or hundres of Mpc. For these conditions, the flavour-transition probability in Eq.~(\ref{EqTransProb}) oscillates very rapidly, and so we are moved to use its average value instead, which is obtained by averaging the oscillating terms in the expression, i.e.
\begin{equation}\label{EqTransProbAvg}
 \langle P_{\alpha\beta} \rangle = \sum_i \lvert U_{\alpha i}\rvert^2 \lvert U_{\beta i}\rvert^2 ~.
\end{equation}
Thus the information in the oscillation phase is lost, but the modifications due to the energy-independent new physics is preserved in the elements of the mixing matrix.

\section{Theoretical framework}\label{Section_Theory}

\subsection{Standard mass-driven flavour oscillations}\label{Section_Theory_Sub_StdOsc}

As we saw in the previous Section, the standard scenario of neutrino oscillations considers each neutrino of a given flavour to be a linear combination of neutrino mass eigenstantes, namely
\begin{equation}
 \lvert \nu_\alpha \rangle = \sum_i \left[U_0\right]_{\alpha i} \lvert \nu_i \rangle ~,
\end{equation}
where we have renamed the mixing matrix $U_0$. The mass eigenstates $\lvert \nu_i \rangle$ satisfy Schr\"{o}dinger's equation and so propagate as
\begin{equation}
 \lvert \nu_i\left(L\right) \rangle = e^{-iHL} \lvert \nu_i \rangle = e^{-i\frac{m_i^2}{2E}L} \lvert \nu_i \rangle ~,
\end{equation}
where we have assumed that $m_i \ll E$, so that $p = \sqrt{E^2-m_i^2} \simeq m_i^2/(2E)$, and that, because neutrinos are highly relativistic particles, $t \simeq L$ (in natural units). Thus, after having propagated for a distance $L$, the neutrino created with flavour $\alpha$ has become
\begin{equation}
 \lvert \nu_\alpha \left(L\right) \rangle = \sum_i \left[U_0\right]_{\alpha i} e^{-i\frac{m_i^2}{2E}L} \lvert \nu_i \rangle ~.
\end{equation}
We see that each mass eigenstate acquires a phase that depends on the value of its mass: the interference of these phases is the source of the flavour transitions. At detection, the probability amplitude of seeing a $\nu_\beta$ is 
\begin{equation}
 \langle \nu_\beta \lvert \nu_\alpha\left(L\right) \rangle 
 = \sum_{i,j} \left[U_0\right]_{\alpha i}^\ast \left[U_0\right]_{\beta j} e^{-i\frac{m_i^2}{2E}L} \langle \nu_j \lvert \nu_i \rangle 
 = \sum_i \left[U_0\right]_{\alpha i}^\ast \left[U_0\right]_{\beta i} e^{-i\frac{m_i^2}{2E}L} 
\end{equation}
and so the probability $P_{\alpha\beta} = \lvert \langle \nu_\beta \lvert \nu_\alpha\left(L\right) \rangle \rvert^2$ for the transition $\nu_\alpha \rightarrow \nu_\beta$ turns out to be Eq.~(\ref{EqTransProb}). (The reader is referred to \cite{PDG08} for a more extended treatment of the basics of neutrino oscillations.)

The oscillation Hamiltonian in the previous description of standard neutrino oscillations was written in the basis of mass eigenstates. An alternative, but equivalent, description of the same phenomenon can be achieved by writing the Hamiltonian in the flavour basis instead. Using a Hamiltonian in the flavour basis will allow us to introduce contributions from new physics in a more straightforward manner. If a neutrino is produced with flavour $\alpha$, then, after having propagated for a distance $L$, its evolved state will be 
\begin{equation}
 \lvert \nu_\alpha\left(L\right) \rangle = e^{-iH_mL} \lvert \nu_\alpha \rangle ~,
\end{equation}
where the oscillation Hamiltonian $H_m$ is the one corresponding to the standard, mass-driven, mechanism, and is written in the flavour basis. $H_m$ is related to the Hamiltonian in the mass basis \--the ``mass matrix''\-- through a similarity transformation that makes use of the unitary mixing matrix $U_0$:
\begin{equation}
 H_m = U_0 H U_0^\dag = U_0 \frac{\text{diag}\left(0,\Delta m_{21}^2,\Delta m_{31}^2\right)}{2E} U_0^\dag ~.
\end{equation}
$U_0$ is the Pontecorvo-Maki-Nakagawa-Sakata mixing matrix, which can be written in terms of three mixing angles, $\theta_{12}$, $\theta_{13}$ and $\theta_{23}$, and one CP-violation phase, $\delta_{CP}$, as
\begin{eqnarray*}\label{EqPMNSMatrix}
 U_0\left(\left\{\theta_{ij}\right\}, \delta_{CP}\right)
 = \left(\begin{array}{ccc}
     c_{12}c_{13} & s_{12}c_{13} & s_{13}e^{-i\delta} \\
     -s_{12}c_{23}-c_{12}s_{23}s_{13}e^{i\delta} & c_{12}c_{23}-s_{12}s_{23}s_{13}e^{i\delta} & s_{23}c_{13} \\
     s_{12}s_{23}-c_{12}c_{23}s_{13}e^{i\delta} & -c_{12}s_{23}-s_{12}c_{23}s_{13}e^{i\delta} & c_{23}c_{13}
 \end{array}\right) ~,
\end{eqnarray*}
with $c_{ij} \equiv \cos\left(\theta_{ij}\right)$, $s_{ij} \equiv \sin\left(\theta_{ij}\right)$.

Using the latest data from solar, atmospheric, reactor (KamLAND and CHOOZ) and accelerator (K2K and MINOS) experiments, the authors of \cite{Schwetz08} found the best-fit values ($1\sigma$) of the standard oscillation parameters to be 
\begin{eqnarray}\label{EqStdMixPar}
 \Delta m_{21}^2 = 7.65^{+0.23}_{-0.20} \times 10^{-5} ~\text{eV}^2 ~~, ~~~ 
 \lvert \Delta m_{31}^2 \rvert = 2.40^{+0.12}_{-0.11} \times 10^{-3} ~\text{eV}^2  \\
 \sin^2\left(\theta_{12}\right) = 0.304^{+0.022}_{-0.016} ~~, ~~~
 \sin^2\left(\theta_{13}\right) = 0.01^{+0.016}_{-0.011} ~~, ~~~
 \sin^2\left(\theta_{23}\right) = 0.50^{+0.07}_{-0.06} ~~.
\end{eqnarray}
The are no experimental values for $\delta_{CP}$ presently.

\subsection{Adding an energy-independent Hamiltonian}\label{Section_Theory_Sub_EnergyInd}

Motivated by the vector coupling considered in Eq.~(\ref{EqLagrangian}), and in analogy to the standard oscillation scenario, we can introduce an energy-independent contribution in the form of the Hamiltonian (also in the flavour basis)
\begin{equation}
 H_b = U_b \text{diag}\left(0,b_{21},b_{31}\right) U_b^\dag ~,
\end{equation}
where $b_{ij} \equiv b_i - b_j$. Following \cite{Dighe08}, we write the mixing matrix in this case as
\begin{equation}
 U_b = \text{diag}\left(0,e^{i\phi_2},e^{i\phi_3}\right)U_0\left(\left\{\theta_{bij}\right\}, \delta_b\right) ~.
\end{equation}
The mixing angles associated with this Hamiltonian are $\theta_{b12}$, $\theta_{b13}$, $\theta_{b23}$, and $\delta_b$ fills the role of $\delta_{CP}$ in the standard Hamiltonian. The two extra phases, $\phi_2$ and $\phi_3$, appear because of how the mixing matrix between flavour and mass eigenstates was defined in Section \ref{Section_Theory_Sub_StdOsc}. 

$H_b$ is dependent on eight parameters \--two eigenvalues ($b_{21}$, $b_{31}$), three mixing angles ($\theta_{b12}$, $\theta_{b13}$, $\theta_{b23}$) and three phases ($\delta_b$, $\phi_2$, $\phi_3$)\-- whose values are currently unknown. Experimental upper limits \cite{Dighe08}, however, have been set for $b_{21}$, using solar and Super-Kamiokande data, and $b_{32}$, using atmospheric and K2K data:
\begin{equation}\label{EqbijLimits}
 b_{21} \le 1.6 \times 10^{-21} ~\text{GeV} ~~, ~~~ b_{32} \le 5.0 \times 10^{-23} ~\text{GeV} ~.
\end{equation}

The full Hamiltonian, including standard oscillations and the energy-independent contribution, is then
\begin{equation}
 H_f = H_m + H_b ~.
\end{equation}
In Section \ref{Section_Intro}, we saw that $H_m$ has been experimentally demonstrated to be the dominant contribution to the oscillations in the low energy (MeV-TeV) regime: according to Eq.~(\ref{EqbijLimits}), the values of the $b_{ij}$ are too low for the new physics, if there is any, to manifest at these energies. The $1/E$ dependence of $H_m$, however, allows us to explore the possibility that, at higher energies, when its contribution is reduced, the effect of an energy-independent Hamiltonian $H_b$ becomes comparable to it or even dominant. Such energy requirement is expected to be fullfilled by the UHE astrophysical neutrino flux (see Section \ref{Section_Intro}.)

We would like to write the flavour transition probability corresponding to this Hamiltonian in a form analogous to Eq. (\ref{EqTransProbAvg}). In order to do this, we need to know what is the mixing matrix $U_f$ between the flavour basis and the basis in which $H_f$ is diagonal. According to basic linear algebra, this is achieved simply by diagonalising $H_f$, finding its normalised eigenvectors, and building $U_f$ by arranging them in column form. The components of the resulting matrix are in general complicated functions of the standard mixing parameters ($\left\{\theta_{ij}\right\}$, $\left\{\Delta m_{ij}^2\right\}$, $\delta_{CP}$) and of the parameters of $H_b$ ($\left\{\theta_{bij}\right\}$, $\left\{b_{ij}\right\}$, $\delta_{b}$, $\phi_2$, $\phi_3$). In analogy to Eq. (\ref{EqTransProbAvg}), the average flavour transition probability associated to the full Hamiltonian $H_f$ is then
\begin{equation}\label{EqTransProbAvgFull}
 \langle P_{\alpha\beta} \rangle = \sum_i \lvert\left[U_f\right]_{\alpha i}\rvert^2 \lvert\left[U_f\right]_{\beta i}\rvert^2 ~.
\end{equation}
By comparing the mixing matrix obtained by diagonalisation of $H_f$ with a general PMNS matrix, Eq.~(\ref{EqPMNSMatrix}) with mixing angles $\Theta_{ij}$ and phase $\delta_f$, we are then able to calculate how the effective mixing angles $\Theta_{ij}$ vary with the parameters of $H_b$ and $\delta_{CP}$. Succintly put, we have
\begin{equation}
 U_f 
 = U_f\left(\left\{\theta_{ij}\right\},\left\{\theta_{bij}\right\},\left\{\Delta  m_{ij}^2\right\},\left\{b_{ij}\right\},\delta_{CP},\delta_{b},\phi_{b2},\phi_{b3}\right) 
 = U_0\left(\left\{\Theta_{ij}\right\},\delta_f\right) ~.
\end{equation}

Up to this point, $U_f$ is dependent on $14$ free parameters. However, the standard mixing parameters $\Delta m_{21}^2$, $\Delta m_{31}^2$, $\theta_{12}$, $\theta_{13}$ and $\theta_{23}$ have been fixed by neutrino oscillation experiments. Additionally, in order to simplify the analysis, in the present work we have set all of the phases to zero, i.e. $\delta_{CP} = \delta_b = \phi_2 = \phi_3 = 0$. Furthermore, also to simplify our analysis, we have made the eigenvalues of $H_m$ proportional to those of $H_b$, at an energy of $E^\star = 1$ PeV, that is,
\begin{equation}
 b_{ij} = \lambda \frac{\Delta m_{ij}^2}{2E^\star} ~,
\end{equation}
with $\lambda$ the proportionality constant. The upper bounds on the $b_{ij}$, Eq.~(\ref{EqbijLimits}), are satisfied for $\lambda \lesssim 10^6$. Thus we are left with only four free parameters to vary: $\lambda$, $\theta_{b12}$, $\theta_{b13}$ and $\theta_{b23}$.

\section{Looking for extreme effects in the flavour ratios}

In the preceding Section, we saw that, in order for the energy-independent contribution to flavour transitions to be visible, we would need to use the expected high-energy astrophysical neutrino flux. As mentioned in Section \ref{Section_Intro}, the sources of this flux, e.g.~active galaxies, are located at distances of tens to hundreds of Mpc, so that the appropriate flavour transition probability to be used is the average one, Eq. (\ref{EqTransProbAvg}). 

If, at the sources, neutrinos of different flavours are produced in the ratios $\phi_e^0:\phi_\mu^0:\phi_\tau^0$, then, because of flavour transitions during propagation, the ratios at detection will be
\begin{equation}\label{EqFluxDetected}
 \phi_\alpha = \sum_{\beta=e,\mu,\tau} \langle P_{\beta\alpha} \rangle \phi_\beta^0 ~,
\end{equation}
for $\alpha = e, \mu, \tau$. Evidently, the initial flavour ratios depend on the astrophysics at source, which is currently not known with high certainty. The standard initial flux \cite{Athar06} considers that the charged pions created in proton-proton and proton-photon collisions decay into neutrinos and muons, which decay into neutrinos, too: 
\begin{equation}
 \pi^+ \rightarrow \mu^+ \nu_\mu \rightarrow e^+ \nu_e \overline{\nu}_\mu \nu_\mu ~~, ~~~
 \pi^- \rightarrow \mu^- \overline{\nu}_\mu \rightarrow e^- \overline{\nu}_e \nu_\mu \overline{\nu}_\mu ~.
\end{equation}
Such process yields $\phi_e^0:\phi_\mu^0:\phi_\tau^0 = 1:2:0$. In the standard oscillation scenario, i.e. in the absence of the energy-independent contribution, plugging this initial flux into Eq.~(\ref{EqFluxDetected}), and using the best-fit values of the mixing angles, Eq.~(\ref{EqStdMixPar}), yields equal detected fluxes of each flavour, i.e. $\phi_e^{std}:\phi_\mu^{std}:\phi_\tau^{std} \approx 1:1:1$ 

In a different production process \cite{Lipari07, Rachen98, Kashti05}, the muons produced by pion decay lose all their energy before decaying, so that a pure-$\nu_\mu$ flux is generated at the source, i.e.~$\phi_e^0:\phi_\mu^0:\phi_\tau^0 = 0:1:0$. In the standard scenario, this initial flux yields a detected flux of $\phi_e^{std}:\phi_\mu^{std}:\phi_\tau^{std} \approx 0.22:0.39:0.39$. Alternatively, a scenario of pure-$\nu_e$ production ($\phi_e^0:\phi_\mu^0:\phi_\tau^0 = 1:0:0$) through beta decay has been considered, e.g.~in \cite{Lipari07}. In this scenario, high-energy nuclei emmited by the source have sufficient energy for photodisintegration to occur, but not enough to reach the threshold for pion photoproduction. The neutrons created in the process generate $\overline{\nu}_e$ through beta decay. For this initial flux, the resulting detected fluxes, in the standard oscillation scenario, are $\phi_e^{std}:\phi_\mu^{std}:\phi_\tau^{std} \approx 0.57:0.215:0.215$. In what follows, we will consider the possibility of observing the energy-independent contribution of $H_b$ assuming these three initial flavour fluxes.

\begin{table}
 \caption{\label{TblStdRS} Standard values (no energy-independent contribution) of the detected fluxes $\phi_\alpha$ ($\alpha = e, \mu, \tau$) and of the ratios $R$, $S$, for the three different initial fluxes considered in the text. The detected fluxes were calculated using the average flavour-transition probability in Eq.~(\ref{EqTransProbAvg}) with the central values of the mixing angles: $\sin^2\left(\theta_{12}\right) = 0.304$, $\sin^2\left(\theta_{13}\right) = 0.01$, $\sin^2\left(\theta_{23}\right) = 0.50$.}
 \begin{center}
  \begin{tabular}{lllll}
   \br
   Production mechanism & Initial flux & Std. detected flux & $R^{std}=\phi_\mu^{std}/\phi_e^{std}$ & $S^{std}=\phi_\tau^{std}/\phi_\mu^{std}$ \\
                        & $\phi_e^0:\phi_\mu^0:\phi_\tau^0$ & $\phi_e^{std}:\phi_\mu^{std}:\phi_\tau^{std}$ &  & \\
   \mr
   Pion decay   & $1:2:0$ & $1:1:1$            & $1$    & $1$ \\
   Muon cooling & $0:1:0$ & $0.22:0.39:0.39$   & $1.77$ & $1$ \\
   Beta decay   & $1:0:0$ & $0.57:0.215:0.215$ & $0.38$ & $1$ \\
   \br
  \end{tabular}
 \end{center}
\end{table}

We have defined the ratios 
\begin{equation}
 R = \frac{\phi_\mu}{\phi_e} ~~, ~~~ S = \frac{\phi_\tau}{\phi_\mu} ~,
\end{equation}
and studied the effects of the new physics on them. The standard values of $R$ and $S$, along with the standard values of the detected fluxes $\phi_\alpha$, for each production scenario, are shown in Table \ref{TblStdRS}. Note that $S=1$ for every production mechanism because the value of $\theta_{23}$ used in the table was its best-fit value $\pi/4$, which ensures yields equal detected fluxes of $\nu_\mu$ and $\nu_\tau$ due to maximal mixing.

\begin{figure}[h!]
  \begin{center}
    \scalebox{0.68}{\includegraphics{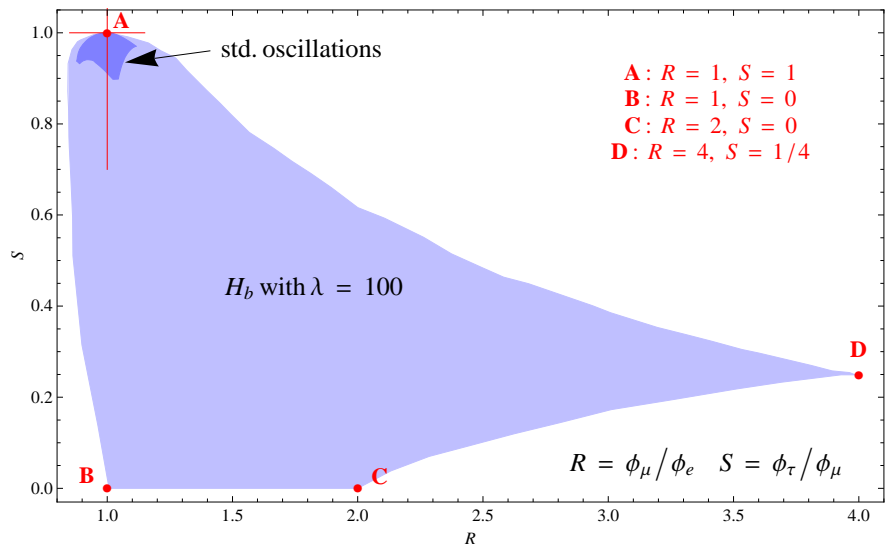}} 
    \caption{Allowed regions of values of $R$ and $S$ assuming an initial flux of $\phi_e^0:\phi_\mu^0:\phi_\tau^0 = 1:2:0$. The light blue region corresponds to the scenario when $\lambda = 100$ (thus making the energy-independent contribution the dominant one) and $\theta_{b12}$, $\theta_{b13}$, $\theta_{b23}$ being allowed to vary, independently, within the range $\left[0,\pi\right]$. The darker-coloured region, shown for comparison, is the allowed region of values obtained when $\lambda = 0$ (that is, energy-independent contributions turned off) and each of the standard mixing angles $\theta_{ij}$ is allowed to vary within its $1\sigma$ bound. A few notable points, marked A-D, are reviewed in Table \ref{TblNotablePts}.}
    \label{FigScatterSR_only120_lambda100}
  \end{center}
\end{figure}

Figure \ref{FigScatterSR_only120_lambda100} shows, in light blue, the allowed region of values of $R$ and $S$ when $\lambda = 100$ and the mixing angles $\theta_{b12}$, $\theta_{b13}$ and $\theta_{b23}$ are allowed to vary, independently, within the range $\left[0,\pi\right]$, with the standard mixing angles set to their best-fit values, Eq.~(\ref{EqStdMixPar}). This value of $\lambda$ already corresponds to the scenario where $H_b$ dominates over $H_m$ in the full Hamiltonian; higher values of $\lambda$ will not modify the shape of the allowed $R-S$ region. For comparison, the darker-coloured region corresponds to the scenario where there is no energy-independent contribution (i.e.~$\lambda = 0$) and the standard mixing angles are allowed to vary within their $1\sigma$ bounds. We have marked four notable points in the plot, A-D. Table \ref{TblNotablePts} shows for each of them the values of the effective mixing angles $\Theta_{ij}$ and the detected fluxes

\begin{figure}[h!]
  \begin{center}
    \scalebox{0.68}{\includegraphics{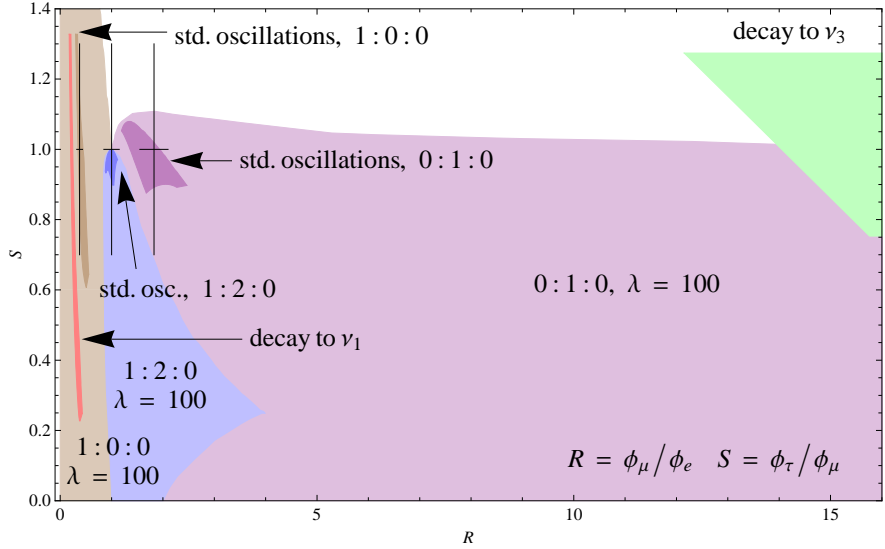}} 
     \caption{Allowed regions of values of $R$ and $S$ for the different initial fluxes $\phi_e^0:\phi_\mu^0:\phi_\tau^0 = 1:2:0$ (blue), $0:1:0$ (purple) and $1:0:0$ (brown). Light blue, light purple and light brown correspond to $\lambda = 100$, a situation where the energy-independent contribution is dominant in the flavour transitions. These regions were generated by varying each $\theta_{bij}$, independently, within $\left[0,\pi\right]$, and fixing the standard mixing angles $\theta_{ij}$ to their best fit values. Dark blue, dark purple and dark brown correspond to $\lambda = 0$, that is, standard oscillations, without any extra contribution. These regions were generated by varying the $\theta_{ij}$ within their $1\sigma$ bounds. Neutrino decay into a single lightest mass eigenstate has been considered, considering both a normal hierarchy (red) and an inverted one (green). These regions were also generated by varying the $\theta_{ij}$ within their $1\sigma$ bounds.}
    \label{FigScatterSR_1sigma_lambda100}
  \end{center}
\end{figure}

If we now consider the two other possible initial fluxes, $\phi_e^0:\phi_\mu^0:\phi_\tau^0 = 0:1:0$ and $1:0:0$, set $\lambda = 100$ to make $H_b$ the dominant contribution and allow each of the $\theta_{bij}$ to vary within $\left[0,\pi\right]$, we obtain the $R-S$ regions in Figure \ref{FigScatterSR_1sigma_lambda100}. The light purple and light brown regions correspond, respectively, to the initial fluxes $0:1:0$ and $1:0:0$ when $\lambda = 100$. The light blue region is the same region that was shown in Figure \ref{FigScatterSR_only120_lambda100}, assuming an initial flux of $1:2:0$ and $\lambda = 100$. The dark blue, dark purple and dark brown regions correpond, respectively, to standard oscillations ($\lambda = 0$), allowing the standard mixing angles $\theta_{ij}$ to vary within their $1\sigma$ bounds. 

\begin{table}[h!]
 \caption{\label{TblNotablePts} Notable points in Figure \ref{FigScatterSR_only120_lambda100}: values of the effective mixing angles $\Theta_{ij}$ and of the detected fluxes.}
 \begin{center}
  \begin{tabular}{cccl}
   \br
   Case & $\left\{\Theta_{12},\Theta_{13},\Theta_{23}\right\}$ & $\phi_e:\phi_\mu:\phi_\tau$ &  \\
   \mr
   A & $\left\{\theta_{12},\theta_{13},\theta_{23}\right\}$            & $1:1:1$ & Standard mixing \\
   B & $\left\{\pi/4,0,0\right\}$     & $1:1:0$ & Maximal mixing $\nu_e\nu_\mu$; $\nu_\tau$'s don't mix \\
   C & $\left\{0,0,0\right\}$         & $1:2:0$ & No effective mixing \\
   D & $\left\{\pi/2,\pi/4,0\right\}$ & $1:4:1$ & Only $\nu_e\nu_\tau$ mix; $\langle P_{e\tau} \rangle = \langle P_{\tau e} \rangle = 1/2$ \\
   \br
  \end{tabular}
 \end{center}
\end{table}

Because we are considering neutrinos that travel distances of tens or hundreds of Mpc, neutrino decay is a possibility. Assuming that the neutrinos decay into products that are not detectable (i.e. ``invisible daughters'' such as sterile neutrinos), then, following \cite{Beacom03}, the flux of flavour $\alpha$ at Earth will be
\begin{equation}
  \phi_\alpha 
  = \sum_{\beta=e,\mu,\tau} \sum_i \phi_\beta^0 \lvert \left[U_0\right]_{\beta i}\rvert^2 \lvert \left[U_0\right]_{\alpha i}\rvert^2 e^{-L/\tau_i}
  ~~\underrightarrow{L\gg\tau_i}~~ \sum_{\beta=e,\mu,\tau} \sum_{i\text{(stable)}} \phi_\beta^0 \lvert \left[U_0\right]_{\beta i}\rvert^2 \lvert \left[U_0\right]_{\alpha i}\rvert^2 ~,
\end{equation}
where $\tau_i$ is the lifetime of the $i$-th mass eigenstate in the laboratory frame. As explained in \cite{Beacom03}, this expression corresponds to the case where the decay has been completed when the neutrinos arrive at Earth.

In a normal hierarchy, $\nu_1$ is the only stable state and so
\begin{equation}
 \phi_\alpha 
 = \lvert \left[U_0\right]_{\alpha1} \rvert^2 \sum_{\beta=e,\mu,\tau} \phi_\beta^0 \lvert \left[U_0\right]_{\beta 1} \rvert^2~,
\end{equation}
while in an inverted hierarchy $\nu_3$ is the stable state and
\begin{equation}
 \phi_\alpha 
 = \lvert \left[U_0\right]_{\alpha3} \rvert^2 \sum_{\beta=e,\mu,\tau} \phi_\beta^0 \lvert \left[U_0\right]_{\beta 3} \rvert^2~,
\end{equation}

In Figure \ref{FigScatterSR_1sigma_lambda100}, decay to $\nu_1$ has been coloured red and decay to $\nu_3$, green. We see that the $R-S$ region for decay into $\nu_1$ does not intersect the region accessible with $\lambda = 100$, assuming an initial flux of $1:2:0$ or $0:1:0$. It is, however, totally contained within the allowed region for $1:0:0$, and thus there is a possibility of not being able to disentangle the effects of neutrino decay (in a normal hierarchy) from an energy-independent contribution, under this production model. The region allowed for decay into $\nu_3$, on the other hand, is superposed to that of $0:1:0$, for values of $R \gtrsim 14$ and $0.75 \lesssim S \lesssim 1.05$. 

A notable feature of this plot is that it reveals that, while there are many opportunities of establishing the presence of an energy-independent contribution by measuring $R$ and $S$, it is a more difficult task to deduce from this measurement what the initial flux was. An extreme example occurs in the region around $R = 2$, $S = 0.2$, where the three assumptions of initial flux result in the same prediction for $R$ and $S$. In other words, if we measured values for $R$ and $S$ around this region, we could conclude (provided there are no other new physics effects) that there is an energy-independent contribution present, but we could not conclude what the initial flux was. On the other hand, if we measured a value of $4 < R < 12$, then (again, assuming there are no unaccounted effects), we could conclude both that there is an energy-independent contribution \textit{and} that the initial flux was $0:1:0$. The same would occur if we measured $R \lesssim 1$, $S \gtrsim 1.35$: in this case, in addition to concluding that an energy-independent contribution is present, we could also conclude that the initial flux was $1:0:0$.

\begin{figure}[h!]
  \begin{center}
    \scalebox{0.68}{\includegraphics{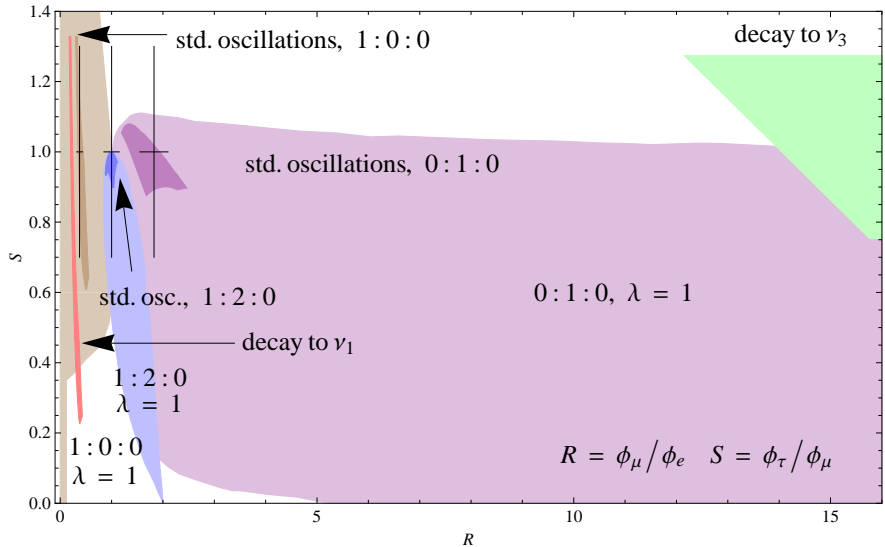}} 
     \caption{Allowed regions of values of $R$ and $S$ for the different initial fluxes $\phi_e^0:\phi_\mu^0:\phi_\tau^0 = 1:2:0$ (blue), $0:1:0$ (purple) and $1:0:0$ (brown). Light blue, light purple and light brown correspond to $\lambda = 1$, a situation where the energy-independent contribution is comparable to the standard terms from mass-driven oscillations. Dark blue, dark purple and dark brown correspond to $\lambda = 0$, that is, standard oscillations, without any extra contribution.}
    \label{FigScatterSR_1sigma_lambda1}
  \end{center}
\end{figure}

As evidenced in Figure \ref{FigScatterSR_1sigma_lambda1}, if we now make the contributions from standard, mass-driven, oscillations and energy-independent new physics comparable, effectively setting $\lambda = 1$, the regions of allowed values of $R$ and $S$ do not change significantly with respect to the case with $\lambda = 100$, corresponding to dominance of the new physics. Hence, it is not necessary for the new physics effects to be dominant in order for them to affect the astrophysical flavour neutrino flux in a measurable way. In the $\lambda = 1$ case, however, we recover some capability to disentangle neutrino decays to $\nu_1$ from the allowed region corresponding to the production mechanism $1:0:0$. Additionally, for values of $S \lesssim 0.1$ and $R \approx 2$, it seems to be possible to identify, in principle, the production mechanism to be $1:2:0$. 

Because $S$ depends on the number of $\nu_\tau$'s detected, and given that this is expected to be a low number (e.g. one every two years at IceCube \cite{Beacom03}), the uncertainty on $S$, using the present generation of neutrino telescopes, will be high. Figures \ref{FigScatterSR_1sigma_lambda1} and \ref{FigScatterSR_1sigma_lambda100} show estimated error bars at the points corresponding to standard oscillations using the best-fit values of the mixing angles, considering a $30\%$ uncertainty on $S$, due to the low statistics of $\nu_\tau$, and a lower value of $15\%$ for $R$, dominated by the systematic error introduced by the atmospheric $\nu_\mu$ flux. Nevertheless, knowledge of only $R$ could be enough to establish whether energy-independent new physics exists at the UHE scale or even to deduce what the initial was. For instance, we have seen that, if a value of $R$ between $4$ and $12$ is measured then we could establish that there is energy-independent new physics and that the initial flux is $0:1:0$.

\section{Summary and conclusions}\label{Section_Discussion}

We have shown that introducing an energy-independent contribution to neutrino flavour oscillations in the form of an additional term in the oscillation Hamiltonian affects the mixing angles, which, in turn, modify the flavour-transition probabilities. This contribution is motivated by scenarios of new physics such as flavour-dependent coupling to the gravitational field and violation of CPT symmetry. Given that the standard oscillation Hamiltonian has a $1/E$ dependence on energy, we have considered in our analysis the expected ultra-high-energy astrophysical neutrino flux, so that it becomes possible for the dominant term in the full oscillation Hamiltonian to be the energy-independent term. This flux, however, propagates over distances of tens or hundres of Mpc, so that we need to use the average flavour-transition probability and, more importantly, it makes sense to consider the possibility that neutrino decays occur. The modifications to the mixing angles result in a change in the expected flux of each flavour that arrives at Earth, which depends both on the initial flux of each flavour and on the oscillation probabilities.

For the initial flavour fluxes, we have considered three possibilities: either neutrinos are produced by the decay of pions generated in proton-proton and proton-photon collisions, and of the muons produced by pion decay, which yields an initial flux of $\phi_e^0:\phi_\mu^0:\phi_\tau^0 = 1:2:0$; or, in a situation where the muons lose most of their energy before decaying, they are produced predominantly as direct products of the decay of pions, which yields a pure-$\nu_\mu$ flux of $0:1:0$; or else they are produced by the beta decay of neutrons liberated in the photodisintegration of nuclei, yielding a pure-$\nu_e$ flux of $1:0:0$. 

We have defined the ratios $R \equiv \phi_\mu/\phi_e$ and $S \equiv \phi_\tau/\phi_\mu$, and shown that large deviations from their standard values are possible, in all three neutrino production scenarios, given our current lack of knowledge about the intervening new physics parameters. Both when the effects of the new physics are dominant as well as when they are comparable to the standard oscillation terms, our ability to deduce, from a single measurement of $R$ and $S$, the existence of an energy-independent contribution \textit{and} the value of the initial flux is dependent on the particular values of $R$, $S$ measured. Given that the number of $\nu_\tau$'s to be detected at current neutrino telescopes is very low (about one every two years at IceCube), however, the uncertainty on $S$ will be quite high. In spite of this, knowledge of only $R$ could be enough to establish the presence of the extra contribution or to deduce what the initial flux was, with particular sensitivity to a $0:1:0$ initial flux.

\ack

This work was supported by grants from the Direccion Academica de Investigacion of the Pontificia Universidad Catolica del Peru (projects DAI-4075 and DAI-L009) and by a High Energy Latinamerican-European Network (HELEN) STT grant. MB acknowledges the hospitality of IFIC during the development of this work.

\end{document}